\documentclass[letterpaper,aps,pre,twocolumn,showcaps,superscriptaddress,amsmath,amssymb]{revtex4}
\usepackage{graphicx}

\newcommand{\mean}[1]{\langle #1\rangle}
\newcommand{\Int}[1]{\int\text{d}#1\;}
\renewcommand{\vec}[1]{\mathbf #1}

\newcommand{\al}{\alpha}
\newcommand{\gam}{\gamma}
\newcommand{\eps}{\varepsilon}
\newcommand{\kap}{\kappa}

\newcommand{\sig}{\sigma}
\newcommand{\ta}{\tau_\alpha}
\newcommand{\tp}{\tobs^\ast}
\newcommand{\tw}{t_\text{w}}
\newcommand{\tobs}{t_\text{obs}}
\newcommand{\cj}{\bar c_\text{in}}
\newcommand{\x}{\vec r}

\newcommand{\ix}{\bar{\vec r}}
\newcommand{\ra}{\rightarrow}
\newcommand{\Nb}{N_\text{b}}

\begin{document}

\title{Constrained dynamics of localized excitations causes a non-equilibrium
  phase transition in an atomistic model of glass formers}

\author{Thomas Speck}
\affiliation{Institut f\"ur Theoretische Physik II,
  Heinrich-Heine-Universit\"at, D-40225 D\"usseldorf, Germany}
\author{David Chandler}
\affiliation{Department of Chemistry, University of California, Berkeley,
  California 94720, USA}

\begin{abstract}
  Dynamical facilitation theory assumes short-ranged dynamical constraints to
  be the essential feature of supercooled liquids and draws much of its
  conclusions from the study of kinetically constrained models. While
  deceptively simple, these models predict the existence of trajectories that
  maintain a high overlap with their initial state over many structural
  relaxation times. We use molecular dynamics simulations combined with
  importance sampling in trajectory space to test this prediction through
  counting long-lived particle displacements. For observation times longer
  than the structural relaxation time exponential tails emerge in the
  probability distribution of this number. Reweighting trajectories towards
  low mobility corresponds to a phase transition into an inactive phase. While
  dynamics in these two phases is drastically different structural measures
  show only slight differences. We discuss the choice of dynamical order
  parameter and give a possible explanation for the microscopic origin of the
  effective dynamical constraints.
\end{abstract}


\maketitle


Since crystallization occurs through nucleation virtually any liquid can be
supercooled below its melting temperature. But some liquids never become
crystals. Their viscosity increases dramatically and at some point internal
relaxation cannot keep up with the cooling anymore and they fall out of
equilibrium, reaching a state we call a \emph{glass} (for reviews see
Refs.~\cite{ange95,debe01}). While in principle protocol-dependent, the
temperature range at which the transition occurs is narrow and nearly a
material property. The idea that this \emph{glass transition} is a, or
determined by a, thermodynamic transition has influenced the theoretical
studies of glasses for decades. However, experimentally determined structure
factors of supercooled liquids show little to no change while approaching the
glass transition. If not global structure, what is the origin of slow
dynamics?

The arguably most striking feature of supercooled liquids is the emergence of
dynamical heterogeneity (see Ref.~\cite{bert10} for a review) below an onset
temperature, i.e., while large regions of the liquid are jammed structural
relaxation continues through regions which are less rigid. This phenomenon has
been observed directly in, e.g., colloidal glasses~\cite{week00} and granular
systems~\cite{keys07}. While simple liquids above the onset temperature are
well described by a mean-field theory~\cite{hansen}, dynamical heterogeneity
leads to different environments for different particles. This observation
forms the foundation for dynamical facilitation theory~\cite{garr03}, a theory
of the glass transition as a dynamical phenomenon, see Ref.~\cite{chan10} for
a review and references. While the interplay between structure and particle
dynamics is complicated, the glass transition is independent of much of these
details and dynamics is dominated by effective constraints restricting the
accessible regions in space-time. The glass transition is controled by the
number of excitations marking locally weak or soft regions able to reorganize,
and not by a thermodynamic variable.

Crucial for dynamical facilitation theory is the notion of a mobility field
coarse-grained both in time and space. A convenient caricature of mobility
fields are spin-like excitations on a lattice~\cite{fred84}. The effect of the
crowded environment on particle motion is mimicked by kinetic constraints,
i.e., for a spin to change its state it must be facilitated by one (or more)
neighboring excited spin(s). These dynamical rules suffice to give rise to
dynamical heterogeneity and a dramatic slow-down of relaxation in the absence
of thermodynamic transitions.

Structural relaxation is the decorrelation of particle positions with their
initial positions over time. In a dense liquid particle motion is strongly
hindered by surrounding particles leading mainly to vibrational motion,
short-lived excursions, and rare, collective, long-lived particle
displacements that could be described as ``cage breaks''~\cite{voll04}. One
approach to glassy dynamics is to find strategies to predict long-time motion
from short-time vibrations probing the local
structure~\cite{widm06,widm09}. In this paper we pursue a different route and
focus on the long-lived displacements as recorders of excitations in the sense
of the simple lattice models. We introduce a binary mobility field but use molecular dynamics to determine its time evolution instead of postulated, and necessarily idealized, dynamics. 

The paper is organized as follows: In Sec.~\ref{sec:kcm} we give a brief
reminder on kinetically constrained models. In Sec.~\ref{sec:exc} we combine
the technique introduced in Ref.~\cite{keys11} to record excitations in
atomistic models of glass formers with the fundamental ideas outlined in
Refs.~\cite{mero05} and \cite{jack06a}. We thus confirm predictions made
previously by the study of kinetically constrained lattice models. In
particular, for observation times much longer than the relaxation time we find
exponential tails in the distribution of mobile particles and a phase
transition upon varying a field that couples to mobility. Such dynamical phase
transitions~\cite{garr09} have been studied analytically and numerically for
kinetic constrained lattice models~\cite{garr07,spec11},
spin-glasses~\cite{jack10}, and have also been found in atomistic glass
formers~\cite{hedg09}. In Sec.~\ref{sec:prop} we then study some structural
and dynamical properties of the inactive phase in more detail before coming to
the conclusions in Sec.~\ref{sec:conc}.

\begin{figure*}
  \centering
  \includegraphics{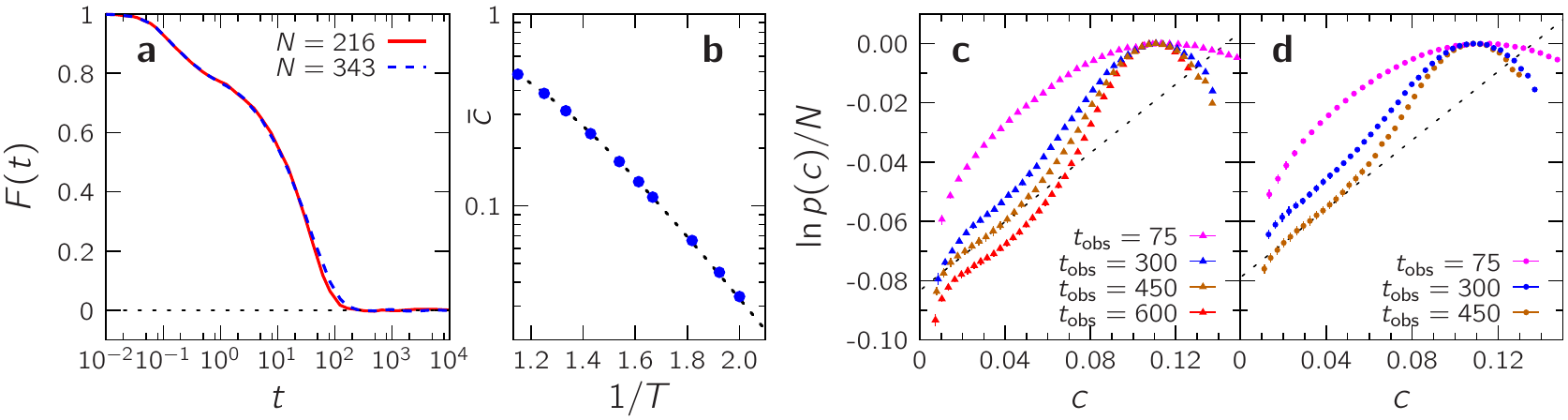}
  \caption{(a)~Intermediate scattering function Eq.~\eqref{eq:isf} at $T=0.6$
    for two system sizes displaying the two-step decay typical for supercooled
    liquids. (b)~The mean density of excited particles $\bar c$ (symbols) as a
    function of inverse temperature $1/T$ below the onset temperature
    $T_0\simeq0.88$. The dashed line is the fit to
    Eq.~\eqref{eq:c}. (c+d)~Logarithm of the probability $p(c)$ of the
    intensive order parameter $c$ scaled by the particle number for (c)
    $N=216$ and (d) $N=343$ particles, and different observation times
    $\tobs$. The dashed lines with slope $\gam=0.58$ indicate the exponential
    tails.}
  \label{fig:exc}
\end{figure*}


\section{Kinetically constrained lattice models}
\label{sec:kcm}

We start with a brief reminder on two popular variants of kinetically
constrained lattice models~\cite{rito03}: the Frederickson-Andersen
(FA)~\cite{fred84} and the East model~\cite{jack91}. Both models have a
trivial energy function $E(\{n_l\})=J\sum_l n_l$ where $J$ sets the energy
scale and $n_l$ is either 1 or 0 indicating whether site $l$ is excited or
not. Excitations correspond to regions of the supercooled liquid where
particles are unjammed and mobile. The effects of local jamming are
incorporated by kinetic constraints: in the FA model a site can change state
only if a neighboring site is excited, i.e., $n_{l\pm1}=1$; in the East model
the site $l+1$ must be excited. Evolution of the mobility field $\{n_l\}$ is
assumed to be Markovian. A simple choice for the rates is the
following. Annihilation $n_l:1\ra 0$ of an excitation occurs with rate 1
defining the time scale. Detailed balance then implies the rate $e^{-J/\tilde
  T}$ for the creation of an excitation (if the constraint is fulfilled). The
equilibrium concentration of excitations is
\begin{equation}
  \label{eq:c}
  \mean{n_l} = (1+e^{J/\tilde T})^{-1}
\end{equation}
with respect to the reduced temperature $1/\tilde T\equiv1/T-1/T_0$. The
temperature $T_0$ marks the onset of dynamical heterogeneity.

As a result of constraints, excitations move in lines.  Fluctuations in
excitations lead to coalescing and branching of excitation lines. These
changes of state, or ``kinks'', provide the means to detect excitations
through particle motion as discussed in the next section. However, the absence
of detectable motion does not signify the absence of excitations which,
without other excitations reaching out, are quiescent as a direct consequence
of the kinetic constraints. The phase transition that we detail in this paper
is a transition between two phases of markedly different concentrations of
kinks or fluctuations.

\section{Particle mobility as space-time order parameter}
\label{sec:exc}


\subsection{Excited particles}

We have performed extensive molecular dynamics simulations on the Kob-Anderson
binary mixture~\cite{kob94}, a popular model for atomistic glass
formers~\cite{yama00,bert10a} (see appendix~\ref{sec:sim} for details). It is
composed of 80\% large (A) and 20\% small (B) particles. Simulations are run
at temperature $T=0.6$ well below the onset temperature $T_0\simeq0.88$ of
heterogeneous dynamics. For comparison, $T\simeq0.435$ is an estimate of the
glass transition temperature for this system~\cite{kob94,bert10a}. We have
chosen the higher temperature and therefore only moderately supercooled state
point to be able to run millions of trajectories over a couple of structural
relaxation times. For that state point, the structural relaxation time is
$\ta\simeq24.5$ as measured by the decay $F(\ta)=1/e$ of the intermediate
scattering function (see Fig.~\ref{fig:exc}a)
\begin{equation}
  \label{eq:isf}
  F(t) = \frac{1}{N}\sum_{l=1}^N \mean{e^{\text{i}\vec
      q\cdot[\x_l(t)-\x_l(0)]}}.
\end{equation}
Particle motion is measured on the length scale $2\pi/|\vec q|\simeq1.058$
corresponding to the peak position of the pair distribution function.
Fig.~\ref{fig:exc}a demonstrates that at $T=0.6$ the structural relaxation for
the system sizes considered here shows no finite size effects (see also
Ref.~\cite{karm09}).

Kinetically constrained models are minimal models incorporating the crucial
ingredient of hindered motion in dense, nearly jammed liquids. What they did
not provide so far is a concrete prescription on how to map a set of particle
positions $\{\x_l\}$ onto a binary mobility field $\{n_l\}$. To establish such
a mapping we follow Ref.~\cite{keys11} and use long-lived particle
displacements of a given length $a$ as recorders of excitations. To this end
we define the single particle indicator function
\begin{equation}
  \label{eq:q}
  h_l(t) \equiv \Theta(|\ix_l(t)-\ix_l(t-\Delta t)|-a),
\end{equation}
where $\Theta(x)$ is the unit step function and $h_l=1$ if the $l$th particle
has moved further than the distance $a$ in the time interval $\Delta t$ and
$h_l=0$ otherwise. In the following we will call particles with $h_l=1$
\emph{mobile}, or \emph{excited} to emphasize their role as recorders of
underlying excitations. To distinguish non-trivial particle displacements from
mere vibrations we use the inherent structure positions
$\{\ix_l\}$~\cite{stil84}. The inherent structure of a configuration is
obtained by steepest descent to the nearest minimum in the potential energy
landscape. Even though there is a hierarchy of motion on different length
scales~\cite{keys11}, here we focus on the single length $a=0.3$. The
commitment time $\Delta t=1.5$ is then chosen to be large enough so that a
particle can commit to a new position but small enough so that we do not count
multiple jumps.

The instantaneous density of mobile particles is
\begin{equation}
  \hat c(t) \equiv \frac{1}{N} \sum_{l=1}^Nh_l(t)
\end{equation}
with mean $\bar c\equiv\mean{\hat c}=\mean{h_l}$. In Fig.~\ref{fig:exc}b we
demonstrate that the temperature dependence of this mean density indeed
follows the prediction for excitations Eq.~\eqref{eq:c}. The fitted onset
temperature $T_0\simeq0.88$ agrees excellently with previous
estimates~\cite{elma09,keys11}. The fitted energy is
$J\simeq3.9$~\footnote{The method employed here to determine mobile particles
  is slightly different from Ref.~\cite{keys11}. As a consequence the energy
  $J_a$ scales differently with length scale $a$. We have checked that
  dynamics is still hierarchical, $J_a-J_1\propto\ln a$, with $J_1\simeq5.4$
  reported in Ref.~\cite{keys11}.}.


\subsection{Low mobility tails}

Our objects of interest are trajectories $X\equiv\{\x_l(t):0\leqslant
t\leqslant\tobs\}$ of fixed length $\tobs$. To quantify the amount of mobility
in a trajectory we count excited particles through the order parameter
\begin{equation}
  \label{eq:op}
  \mathcal C[X] \equiv \sum_{i=1}^K \sum_{l=1}^N h_l(t_i),
  \qquad
  c \equiv \mathcal C/(NK)
\end{equation}
with equally spaced $t_i=i\Delta t$ and observation time $\tobs=K\Delta t$.
In Fig.~\ref{fig:exc} we show distributions $p(c)$ for different system sizes
and different observation times obtained through umbrella sampling combined
with replica exchange (see appendix~\ref{sec:umbrella}). For trajectories in
which motion decorrelates on a time scale much shorter than $\tobs$ we would
sum over many independent events. Following the central limit theorem the
probability distribution $p(c)$ then approaches a Gaussian. Indeed, for
moderate observation times $\ta<\tobs\lesssim\tp$ larger than the relaxation
time $\ta$ but below a cross-over time $\tp$ we find such Gaussian
distributions (see $\tobs=75\approx3\ta$ in Fig.~\ref{fig:exc}). Increasing
$\tobs\gtrsim\tp$ we observe two effects: for small $c$ exponential tails
emerge and the shape of $p(c)$ becomes non-concave. The physical picture is
that highly constrained dynamics facilitates the creation of extended immobile
regions, i.e., compared to uncorrelated dynamics it is easier to ``remove''
mobility from a given space-time volume.

The explanation why the tails of $p(c)$ are exponential is as
follows~\cite{mero05}. Assuming that excitations are non-interacting the
probability to find at $t=0$ an immobile region of $\ell$ particles is
proportional to $e^{-\gam\bar c\ell}$ with a geometric factor $\gam$
independent of $N$ and $\tobs$. Combining this probability with $\mathcal
C\approx K(N-\ell)\bar c$ for the case that this ``bubble'' persists leads to
$\ln p(c)=\gam Nc$ plus an offset. In Ref.~\cite{keys11}, evidence is
presented that the assumption of non-interacting excitations is indeed a good
approximation. Of course, not all bubbles span the entire trajectory
connecting the initial with the final state. The temporal extent $\tp$ of a
typical bubble grows proportionally to the mean persistence time, i.e., the
mean time a particle remains at its initial inherent structure position.

The order parameter Eq.~\eqref{eq:op} is purely dynamical. Since in a crystal
particle mobility is low and motion restricted to defects such an order
parameter cannot discriminate between a low activity amorphous and a low
activity crystalline phase. Therefore, we also monitor the structure through
the orientational order parameter $\psi_6(l)$ defined in
Eq.~\eqref{eq:psi6}. To allow local fluctuations in structure but prevent
global long-range order we use the value of
\begin{equation}
  \bar\psi_6 \equiv \frac{1}{N_\text{A}}\sum_{l=1}^{N_\text{A}} \psi_6(l)
\end{equation}
averaged over the $N_\text{A}=0.8N$ large particles of the final configuration
of trajectories. We reject all trajectories with $\bar\psi_6>0.45$. (see also
Fig.~\ref{fig:struc}c below)


\subsection{Active--inactive phase transition}

So far we have established that trajectories contributing to the exponential
tails in Fig.~\ref{fig:exc} are those which remember their initial conditions
and do not relax within the observation time $\tobs$. At least formally and in
computer simulations we can apply a bias in the space of trajectories to
stabilize these low mobility trajectories. The fact that the distributions
$p(c)$ are non-concave implies a phase transition between an active phase
corresponding to the liquid in which motion is plentiful, and an inactive
phase of low particle mobility. To provide a link with traditional
thermodynamics, and the Ising model in particular, imagine the $\{h_i\}$ to be
spins on a lattice with the order parameter $\mathcal C$ taking the role of
the magnetization~\cite{chandler}. Below the critical temperature the system
undergoes a first-order transition between a disordered phase of low
magnetization and an ordered phase of high magnetization through applying a
field (which we will call $s$ in the following). While the statistical
treatment is analogous, the underlying physics of a supercooled liquid is of
course different from the Ising model. In our case the lattice extends over
space \emph{and} time, and the interactions between ``spins'' is due to
short-ranged forces, geometrical confinement, and the thereof resulting
temporal correlations of particle motion.

\begin{figure}[t]
  \centering
  \includegraphics{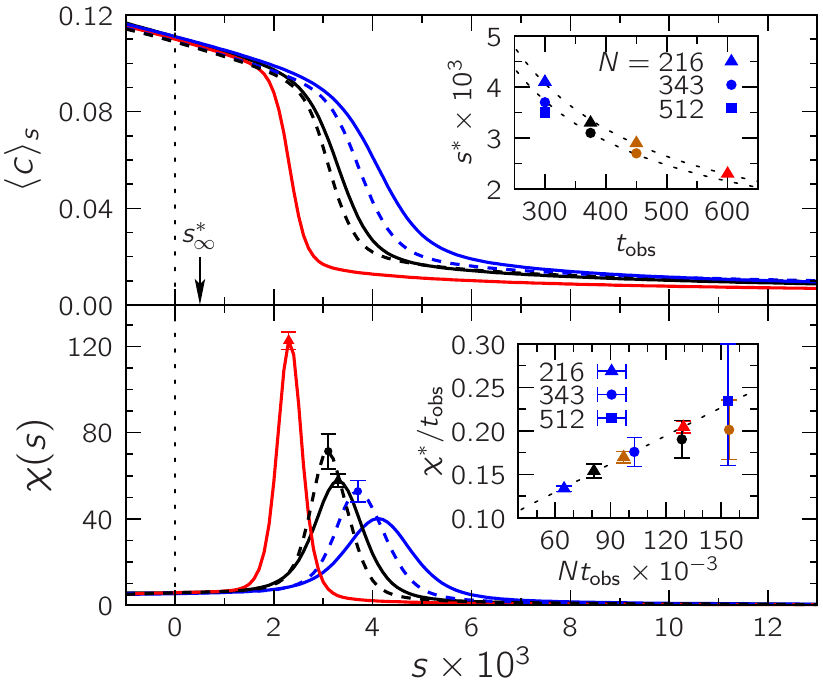}
  \caption{Mean fraction of excited particles (top) and susceptibility
    (bottom) \textit{vs}. the biasing field $s$ for selected observation times
    (from right to left: $\tobs=300,375,600$) and system sizes $N=216$ (solid
    lines) and $N=343$ (dashed lines). For clarity we only show error bars for
    the peak values of the susceptibility $\chi(s)$. Upper inset: Dependence
    of the coexistence field $s^\ast$ on trajectory length $\tobs$ and system
    size $N$. The dashed lines are fits to $s^\ast=s^\ast_N+a/\tobs$. The fit
    parameters $s^\ast_N$ are $s_{216}\simeq5.1\times10^{-4}$ and
    $s_{343}\simeq5.8\times10^{-4}$. From these results we estimate the
    coexistence field to be $s^\ast_\infty\approx5\times 10^{-4}$ (arrow) in
    the limit $\tobs\ra\infty$. Lower inset: System size dependence of
    $\chi^\ast$.}
  \label{fig:phase}
\end{figure}

In Fig.~\ref{fig:phase} we plot the mean fraction of excited particles
\begin{equation}
  \label{eq:cs}
  \mean{c}_s \equiv \frac{\mean{c e^{-s\mathcal C}}}{\mean{e^{-s\mathcal C}}}
  = \frac{\Int{c} c p(c)e^{-sNKc}}{\Int{c} p(c)e^{-sNKc}}
\end{equation}
and the susceptibility $\chi(s)\equiv-\partial\mean{c}_s/\partial s$
\textit{vs.} the biasing field $s$. The plot shows that the density of mobile
particles drops from $\bar c\simeq0.11$ at $s=0$ to about $\cj\simeq0.01$ for
$s\gg s^\ast$. For small $s$ we can expand the mean $\mean{c}_s\approx\bar
c-\kap s$ with $\kap\equiv[\mean{C^2}-\mean{C}^2]/(NK)$. The linear behavior
around $s=0$ in Fig.~\ref{fig:phase}, therefore, reflects the Gaussian nature
of the liquid phase. The coexistence field $s^\ast$ is obtained from the peak
position of $\chi(s)$ maximizing the fluctuations of the order
parameter. Increasing either the number of particles $N$ or the observation
time $\tobs$ sharpens the transition. However, space and time are not
symmetric. At least in part, this asymmetry reflects that we employ periodic
boundary conditions in space whereas trajectories can have quite distinct
initial and final states. This leads to temporal boundary effects enhancing
the mobility at the beginning and the end of the trajectory~\cite{garr09}. For
fixed $N$ one expects to leading order $s^\ast=s^\ast_N+\mathcal O(1/\tobs)$
as shown in the upper inset of Fig.~\ref{fig:phase}. From the fits we estimate
the limiting coexistence field $s^\ast_\infty\approx5\times10^{-4}$ to be
small but nonzero. Finally, in the lower inset of Fig.~\ref{fig:phase} we
demonstrate the finite-size scaling of the peak values
$\chi^\ast\equiv\chi(s^\ast)$ plotted \textit{vs}. the space-time volume
$N\tobs=NK\Delta t$. The dashed line corresponds to
$\chi^\ast\approx\tobs(0.065+1.08\times10^{-6}N\tobs)$, which suggests a
first-order transition with a diverging susceptibility and a discontinuous
jump of $\mean{c}_s$ at $s^\ast_\infty$ in the limit of large $N$ and/or
$\tobs$.


\subsection{Choice of order parameter}

\begin{figure}[b!]
  \centering
  \includegraphics{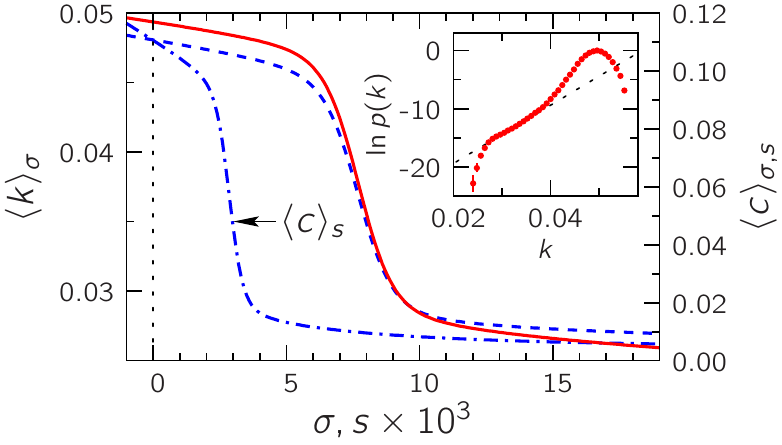}
  \caption{Mean intensive activity $\mean{k}_\sig$ \textit{vs.} the biasing
    field $\sig$ for $N=216$ and $\tobs=450$ (solid line). For comparision,
    the mean fraction of excited particles (right axis) is shown for both the
    ensembles defined through $\mathcal K$ ($\mean{c}_\sig$, dashed line) and
    $\mathcal C$ ($\mean{c}_s$, dash-dotted line). Inset: Probability
    distribution of $k$. The dashed line indicates the exponential tail.}
  \label{fig:activity}
\end{figure}

To check whether the transition into an inactive phase is robust with respect
to the way we measure mobility in trajectories, we have also considered the
dynamical order parameter
\begin{equation}
  \label{eq:K}
  \mathcal K[X] \equiv \sum_{i=1}^K\sum_{l=1}^N |\x_l(t_i)-\x_l(t_{i-1})|^2,
  \quad
  k \equiv \mathcal K/(N\tobs)\,.
\end{equation}
This order parameter was used in Ref.~\cite{hedg09} to demonstrate for the
first time a transition between a high activity and a low activity phase in an
atomistic model. It sums over the short-time mean-square displacements of
particles, which obscures the separation of vibrations from reorganization
events that lead to structural relaxation. Nevertheless, as demonstrated in
Fig.~\ref{fig:activity}, an abrupt transition from high to low activity can
still be observed. To be specific, we define a new ensemble
\begin{equation}
  \mean{A}_\sig \equiv
  \frac{\mean{A e^{-\sig\mathcal K}}}{\mean{e^{-\sig\mathcal K}}},
\end{equation}
where we denote the biasing field coupling to $\mathcal K$ by $\sig$ and $A$
is any observable. As an illustration for $N=216$ and $\tobs=450$ the mean
$\mean{k}_\sig$ is plotted in Fig.~\ref{fig:activity}. It resembles the curves
shown in Fig.~\ref{fig:phase} and drops abruptly around
$\sig^\ast\simeq0.0077$. Moreover, we find exponential tails in the
probability distribution of $k$ plotted in the inset of
Fig.~\ref{fig:activity}. In addition to $\mean{k}_\sig$ we determine the
density of mobile particles $\mean{c}_\sig$, which closely follows the
former. However, while the activity drops by a factor of less than two, the
number of excited particles is reduced by more than one order of
magnitude. The comparision with the curve $\mean{c}_s$ obtained in the
$\mathcal C$-ensemble shows that the mean fraction of mobile particles
approaches the same value (within uncertainties) in both ensembles. We,
therefore, conclude that both measures $\mathcal C$ and $\mathcal K$ prepare
the same inactive phase through applying a biasing field in trajectory
space. The transition is more pronounced in $\mathcal C$ since vibrational
motion, which does not cease in the inactive phase evolving at the same
temperature as the active phase, contributes significantly to
Eq.~\eqref{eq:K}.

\section{Properties of the inactive phase}
\label{sec:prop}


\subsection{Nucleation of activity}

An important consequence of first order phase transitions is nucleation: a
system crossing the transition line, although the new phase is favored, has to
pay a penalty for interfaces and one has to wait for a large enough nucleus to
appear spontaneously. Translated to the present case one might ask what
happens if at time $t=\tobs$ for $s\gg s^\ast$ we turn off the field $s$. In
the picture of facilitated dynamics a new excitation can only appear close to
an existing excitation. Since both density of excitations and kinks, as
recorded through $\mean{c}_s$, are drastically reduced in the inactive phase
we have to wait a certain time before excitations ``percolate'' through the
system and it returns to the liquid phase. The nucleation of activity,
therefore, is conceptually different from, say, the nucleation of a crystal
which is determined by a single large barrier.

\begin{figure}
  \centering
  \includegraphics{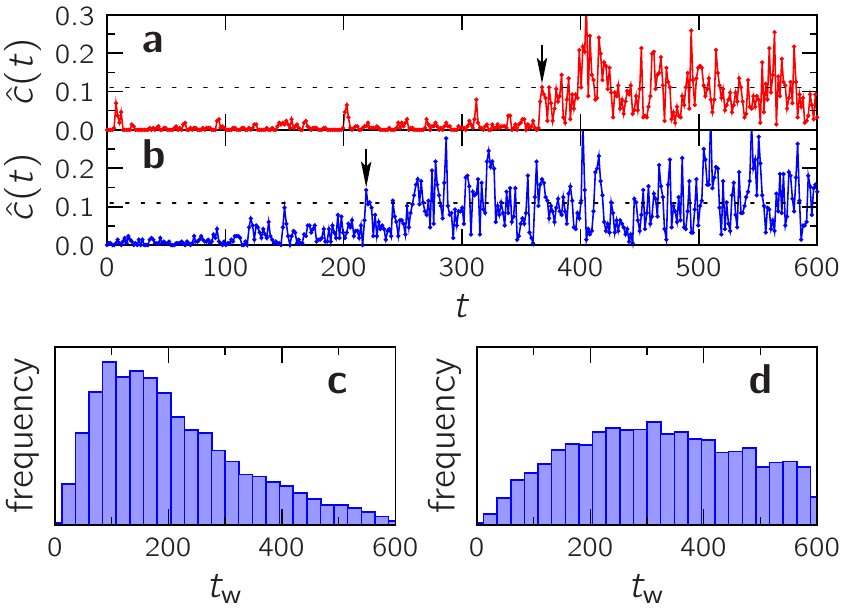}
  \caption{Melting of the inactive phase: (a)~Fraction of mobile particles
    $\hat c(t)$ \textit{vs}. time for an initial configuration prepared with
    the $s$-ensemble at $s=0.01$. The systems remains inactive up to time
    $\tw$ (arrow), when it suddenly unjams. The dashed line indicates the
    average density $\bar c$. (b)~Another melting trajectory showing a more
    gradual ``thawing''. (c)~The distribution of $\tw$ for 10,000 trajectories
    starting out of the same initial configuration. (d)~A different initial
    configuration showing a much broader distribution of waiting times.}
  \label{fig:nuc}
\end{figure}

To demonstrate this ``melting'' of jammed configurations we prepare
trajectories at $s=0.01$. From these trajectories we then take a single
configuration, randomize velocities, and run 10,000 unbiased trajectories out
of this initial configuration (see also the isoconfigurational
ensemble~\cite{widm04}). In Fig.~\ref{fig:nuc}a a single trajectory is
shown. Clearly, the system remains inactive for many structural relaxation
times (in this example $\tw\simeq15\ta$) as measured by $\hat c$ and then,
suddenly, becomes active again with large fluctuations of $\hat c$. In
Fig.~\ref{fig:nuc}b a different trajectory out of the same initial
configuration shows a more gradual transition from inactive to active. In
Fig.~\ref{fig:nuc}c and d we show distributions of waiting times $\tw$ for two
different initial states. These distributions are clearly non-exponential,
which is consistent with a variable step process as excitations reach out and
reconnect. A detailed comparision of these distributions to predictions from
kinetically constrained models is left for a future study.


\subsection{Local structure}

\begin{figure*}
  \centering
  \includegraphics{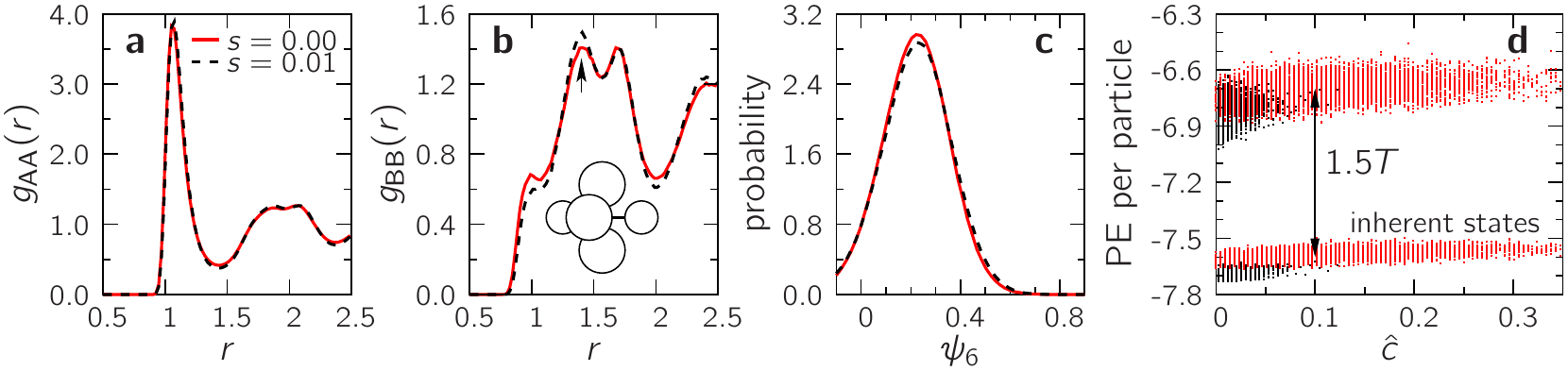}
  \caption{Different measures of the structure in the liquid ($s=0$) and the
    inactive phase ($s=0.01$) for $N=216$ and $\tobs=600$: (a)~Radial
    distribution function for the large A particles and (b)~for the small B
    particles. The arrow marks the peak corresponding to a B-B bond with 3
    common A neighbors, see sketch and main text. (c)~Distribution of the
    structural order parameter $\psi_6$ measuring long range order. A particle
    in a perfect crystal would have $\psi_6=1$. (d)~Scatter plot of potential
    energy (PE) per particle versus the concentration of excited particles
    $\hat c$ for both actual positions and inherent states. Red points are
    from the ensemble of active states and black points are from the ensemble
    of inactive states. The harmonic contribution $3T/2$ to the potential
    energy is indicated.}
  \label{fig:struc}
\end{figure*}

In the introduction we have emphasized that global structural differences
between liquid and glass are at most minuscule. However, the fact that the
system remains jammed for particle configurations taken from trajectories
prepared at $s\gg s^\ast$ indicates that there is a structural difference
between these configurations and configurations typically visited in the
liquid phase. To make this more quantitative we sample trajectories at fixed
$s$ and compare the structures as measured by three different methods, see
Fig.~\ref{fig:struc}. The pair distribution function $g_\text{AA}(r)$ for the
large (A) particles shown in Fig.~\ref{fig:struc}a demonstrates that liquid
and inactive phase are globally indistinguishable. Small differences are seen
for the small (B) particles in Fig.~\ref{fig:struc}b. Beyond the simple
two-point functions we also consider the histogram of the bond-order parameter
$\psi_6$ as defined in appendix~\ref{sec:order} plotted in
Fig.~\ref{fig:struc}c. This order parameter is a convenient measure for
long-range order. For every particle it quantifies its local order with
$\psi_6=1$ for a particle in a perfect crystal. All measures clearly show that
the inactive phase is amorphous.

In Fig.~\ref{fig:struc}d we show that the potential energy per particle and
the density of mobile particles are uncorrelated in both phases. While in the
inactive phase the potential energy of particles is typically lower, the mean
difference $\approx0.1$ is much less than the vibrational contribution
$\approx1.5T$ separating real space potential energies from the inherent state
energies. Moreover, as demonstrated in Fig.~\ref{fig:struc}d, there is still
an overlap of potential energies between both phases. Hence, we conclude that
particles are not trapped energetically but rather due to geometrical
constraints.

Differences in structure are picked up by the pair distribution function for
the small (B) particles plotted in Fig.~\ref{fig:struc}b. It has been shown
that the peaks of $g_\text{BB}(r)$ for binary mixtures can be assigned to
certain local structures: two bonded B particles sharing $m$ common A
neighbors~\cite{fern04}. Of particular interest is the second peak
corresponding to $m=3$ since it indicates icosahedral coordination
shells. Fig.~\ref{fig:struc}b shows that in the inactive phase this local
structure occurs more often compared to $s=0$. This is consistent with recent
observations of short-ranged structures in supercooled binary
mixtures~\cite{cosl07,pede10}. Slow relaxation is attributed to reorganization
of particles bound in these structures. Moreover, the drop of $\approx0.1$ in
the potential energy of inherent states agrees quantitatively with the drop
associated to the formation of these structures~\cite{pede10} (albeit for a
slightly different model).


\subsection{Dynamical facilitation}

\begin{figure}[b!]
  \centering
  \includegraphics{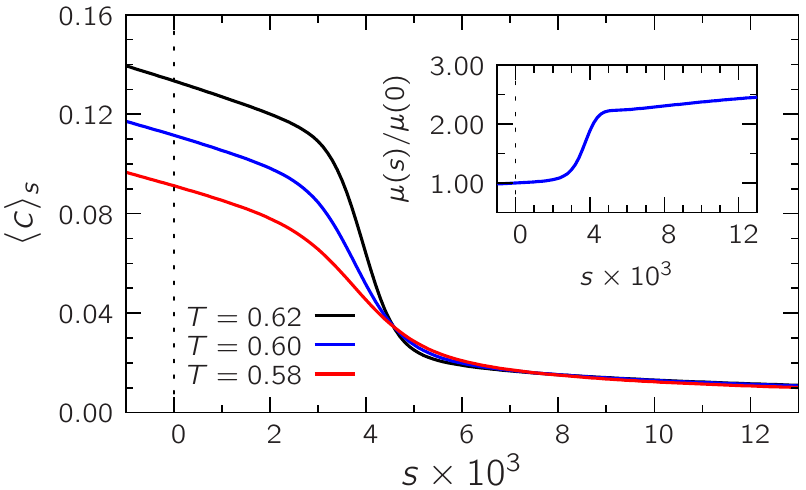}
  \caption{Mean fraction of excited particles $\mean{c}_s$ at three different
    temperatures $T$. Inset: The normalized transfer function
    Eq.~\eqref{eq:mu} showing an increase of facilitation at large $s$. (all
    data for $N=216$ and $\tobs=300$)}
  \label{fig:dynfac}
\end{figure}

We finally study the behavior of facilitation when going from the active to
the inactive phase. First, we note that the fraction of excited particles as
plotted in Fig.~\ref{fig:dynfac} is independent of temperature in the inactive
phase. This indicates that the dynamics in the inactive phase is decoupled
from the externally fixed temperature. Second, we study the degree to which
particle motion is facilitated. Different methods have been reported in the
literature including a mobility transfer function~\cite{voge04} and the
facilitation volume~\cite{keys11}. In the spirit of a mobility transfer we
consider the set of newly excited particles for which the binary indicator
function
\begin{equation}
  w_l(t) \equiv [1-h_l(t-\Delta t)]h_l(t)
\end{equation}
is $w_l=1$. We follow a single particle along a trajectory and through
\begin{equation}
  \mathcal N[X] \equiv \sum_{i=0}^{K-1}\sum_{l=1}^N
  h_1(t_i)w_l(t_{i+1})\Theta(r-|\ix_1(t_i)-\ix_l(t_i)|)
\end{equation}
we count the number of excited particles that have been created in a sphere
with radius $r$ around the tagged particle under the condition that the tagged
particle itself had been excited in the preceding time slice. We define the
transfer function
\begin{equation}
  \label{eq:mu}
  \mu(s) \equiv \frac{\mean{\mathcal N}_s}{K\mean{c}_s\mean{w}_s}.
\end{equation}
The ratio $\mu(s)/\mu(0)$ is plotted in the inset of Fig.~\ref{fig:dynfac}
using $r=1.5$ roughly corresponding to the first coordination shell. It shows
that the probability that a particle becomes excited close to an already
mobile particle increases in the inactive phase.

Putting all observations together the following picture of dynamics in the
inactive phase emerges: The persistence time exceeds the observation time and
most particles maintain a high overlap with their initial position. However,
some activity continues in isolated regions. The fraction of these mobile
particles is decoupled from the temperature. Particles do not become mobile
(or immobile) at random but are facilitated through existing mobile particles
in their vicinity. Turning off the $s$-field these remaining mobile particles
are the seeds from which excitations can reconnect before the system returns
to its fluid state.


\section{Conclusions}
\label{sec:conc}

The separation between fast inter-basin vibrations and slow, activated
transitions between inherent structures (or meta-basins~\cite{heue08}) is the
essence of the energy landscape paradigm~\cite{gold69,debe01}. It implies a
time evolution that is dominated by rare thermal fluctuations that carry the
system from one minimum over a barrier into a neighboring
minium~\cite{fris11}. However, there is mounting evidence that such a
mechanism competes with, or is even shadowed by, relaxation that occurs
through channels that present only low energetic barriers but which are rare
and found through ``surging'' particle motion: examples are string-like
motion~\cite{gebr04} and participation maps of low-frequency
modes~\cite{widm04,chen11}. Dynamical facilitation theory postulates
``excitations'' to be the fundamental objects describing such structural
weaknesses, and facilitation to be the dominant mechanism at low
concentrations of excitations.

The energy landscape picture assumes that the way the landscape is sampled is
governed by temperature. From the evidence presented here, we arrive at a
seemingly different perspective: the concentration of excitations determines
relaxation. At constant temperature the system can be forced into an inactive
\emph{glassy} phase through either removing excitations or suppressing
fluctuations leading to ``frozen'' excitations that are quiescent. In this
inactive phase particles vibrate around local energy minima, the statistics of
which is consistent with the externally fixed temperature. In contrast,
transitions between local minima, or inherent states, are rare and decoupled
from this temperature. It appears that these jammed states, in which
excitations are arrested, can be created rather easily through local particle
rearrangements that do not affect the global structure. In Ref.~\cite{jack11}
it is shown that these states are mechanically more stable than fluid states
at a lower temperature. Here we have demonstrated that the melting of jammed
states is not consistent with a single crossing of a large free energy barrier
but that it is rather a multi-step process. We attribute this multi-step
process to the ``unfreezing'' of excitations, an interpretation that is
supported by an increased degree of facilitation in these jammed states. These
observations, together with the emergence of exponential tails equivalent to
those observed in kinetically constrained models, leads us to the conclusion
that the transition is indeed caused by local dynamic constraints. The precise
pathways and microscopic mechanisms of the particle rearrangements underlying
the active-inactive transition are left for future studies.

As a final note we emphasize that the active to inactive transition is
reminiscent of the glass transition. The fundamental difference is that the
transition demonstrated in this paper is a transition controlled by a field
coupling to a dynamical observable, while the experimental glass transition is
controlled by rate of temperature decrease. The connection between the two
remains to be quantified.


\acknowledgments

We have profited from discussions with A.S. Keys and U.R. Pedersen. We thank
L.O. Hedges for assistence with early stages of the code and J.D. Chodera for
help in understanding and implementing MBAR. TS was supported in part by the
Alexander-von-Humboldt foundation, and TS and DC were supported in part by the
Director, Office of Science, Office of Basic Energy Sciences, Materials
Sciences and Engineering Division and Chemical Sciences, Geosciences, and
Biosciences Division of the U.S. Department of Energy under Contract
No.~DE-AC02-05CH11231.


\appendix

\section{Simulation details}
\label{sec:sim}

We have performed extensive molecular dynamics simulations on the 
binary mixture~\cite{kob94}. It is composed of 80\% large (A) and 20\% small
(B) particles possessing the same mass $m$. Particles interact through the
continuous truncated and shifted Lennard-Jones potentials
\begin{equation*}
  u_{\al\beta}(r) =
  \begin{cases}
    u_\text{LJ}(r;\eps_{\al\beta},\sig_{\al\beta}) - \\
    u_\text{LJ}(2.5\sig_{\al\beta};\eps_{\al\beta},\sig_{\al\beta}), &
    r \leqslant 2.5\sig_{\al\beta} \\
    0, & r > 2.5\sig_{\al\beta}
  \end{cases}
\end{equation*}
where $u_\text{LJ}(r;\eps,\sig)=4\eps[(\sig/r)^{12}-(\sig/r)^6]$. The
parameters read $\sig_\text{AA}=\sig$, $\sig_\text{AB}=0.8\sig$,
$\sig_\text{BB}=0.88\sig$, $\eps_\text{AA}=\eps$, $\eps_\text{AB}=1.5\eps$,
and $\eps_\text{BB}=0.5\eps$. Simulations are run at constant volume $V$,
constant temperature $T$, and constant number density $N/V=1.2$ with $N$ the
number of particles at positions $\{\x_l\}$. Throughout the paper we employ
reduced Lennard-Jones units with respect to the large particles, i.e., we
measure length in units of $\sig$, energy in units of $\eps$, time in
units of $\sqrt{m\sig^2/\eps}$, and we set Boltzmann's constant to unity.

Newton's equations of motion are integrated through the velocity Verlet
algorithm with time step 0.005 using LAMMPS~\cite{lammps}. For energy
minimization we employ the FIRE algorithm~\cite{bitz06}.

\section{Importance sampling}
\label{sec:umbrella}

Just as standard Monte Carlo simulations~\cite{frenkel} do importance sampling
of configurations, the methods we employ do importance sampling of
trajectories. Specifically, we harvest new trajectories using moves from
transition path sampling~\cite{bolh02,dell02}. These moves preserve the
equilibrium weight $P_0[X]$ of trajectories. Hence, accepting or rejecting a
move $X\ra X'$ according to the usual Metropolis criterion
\begin{equation*}
  \min \left\{ 1,\frac{e^{-w(\mathcal C[X'])}}{e^{-w(\mathcal C[X])}} \right\}
\end{equation*}
generates an ensemble of trajectories with weight
\begin{equation*}
  P[X] \sim P_0[X] e^{-w(\mathcal C[X])}.
\end{equation*}
Here, $\mathcal C[X]$ is a dynamical order parameter that calculates a real
number from trajectory $X$, see Eqs.~\eqref{eq:op} and~\eqref{eq:K}; and
$w(x)$ is the weight function.

For a single trajectory we store $K+1$ configurations at times $t_i=i\Delta t$
with $i=0,\dots,K$. We employ the ``massive stochastic collision'' thermostat,
i.e., all velocities are randomized at these times and the center-of-mass
velocity is subtracted. The use of a stochastic thermostat allows us to
perform transition path sampling with so-called 'half moves' as described in
detail in Ref.~\cite{dell02}. In order to efficiently sample probability
distributions of the order parameter we use the quadratic form
\begin{equation}
  \label{eq:w}
  w(x) = \frac{k_\text{s}}{2}(x-x_0)^2.
\end{equation}
To speed up the sampling of trajectory space and decrease the correlations of
subsequently generated trajectories we use replica exchange between
$N_\text{rep}=8$ replicas with different values of $\{x_0\}$. Hence, a single
cycle consists of two consecutive steps: (i) every replica generates a new
trajectory which is either accepted (and replaces the previous trajectory) or
rejected, and (ii) trajectories are swapped between replicas. To obtain good
mixing we attempt $8^5$ swaps between all of the replicas, not only neighbors.
Data has been acquired from two independent runs, i.e., two independent seed
trajectories (except $K=50$, which is from one run). For a single run we let
the trajectories relax for $N_\text{x}$ cycles and then recorded $N_\text{t}$
trajectories. Table~\ref{tab:ov} shows an overview for the data gathered to
produce Figs.~\ref{fig:exc} and~\ref{fig:phase}.

\begin{table}
  \centering
  \caption{System sizes studied and number of harvested trajectories.}
  \label{tab:ov}
  \begin{tabular*}{\hsize}{@{\extracolsep{\fill}}cccc}
    system size & trajectory length & $N_\text{x}$ & total $N_\text{t}$ per
    replica \\
    \hline
    $N=216$ & $K=50$ & 2000 & 50000 \\
    & $K=200$ & 2000 & 40000 \\
    & $K=250$ & 2000 & 40000 \\
    & $K=300$ & 2000 & 50000 \\
    & $K=400$ & 2000 & 50000 \\
    \hline
    $N=343$ & $K=50$ & 2000 & 50000 \\
    & $K=200$ & 2000 & 50000 \\
    & $K=250$ & 3000 & 60000 \\
    & $K=300$ & 6000 & 60000 \\
    \hline
    $N=512$\tablenote{Low activity umbrellas might not have fully
      equilibrated.} & $K=200$ & 10000 & 30000 \\
    \hline
  \end{tabular*}
\end{table}

\section{MBAR}
\label{sec:mbar}

To calculate distributions and expectation values from raw data we use Shirts'
and Chodera's multistate Bennett acceptance ratio (MBAR) method~\cite{shir08}
and its extension to path ensembles~\cite{minh09}. For completeness we briefly
summarize the method. We solve
\begin{equation*}
  f_i = -\ln \sum_{j=1}^{N_\text{rep}}\sum_{n=1}^{N_\text{t}}
  \frac{e^{-w_i(x_{jn})}}{N_\text{t}\sum_{k=1}^{N_\text{rep}}(g_j/g_k)
    e^{f_k-w_k(x_{jn})}}
\end{equation*}
self-consistently for the set of ``free energies'' $\{f_i\}$. Here, $w_i(x)$
is the weight function corresponding to replica $i$, and $x_{jn}=w_j(\mathcal
C[X_n])$ is the value of the order parameter along the $n$th trajectory of
replica $j$. While in principle MBAR provides an error estimate it requires
independent samples, whereas consecutive trajectories obtained using the
method described in the previous section are highly correlated. The
statistical inefficiency of replica $j$ is $g_j=1+2\tau_j$, where $\tau_j$ is
the correlation time (in unit samples) of some representative observable (here
we use $c$). One possibility to obtain independent samples is to subsample the
data series with stride $g_j$. Here we use all data but weigh replicas
according to their relative statistical inefficiencies. Errors are estimated
by splitting the data into chunks of $N_\text{t}=10,000$ trajectories and
calculating the standard error of expectation values.

Expectation values of an observable $A$ are calculated through
\begin{equation*}
  \mean{A} \simeq \sum_{j=1}^{N_\text{rep}}\sum_{n=1}^{N_\text{t}} A_{jn}
  e^{-[w(x_{jn})-f]}.
\end{equation*}
Here, $w(x)$ and $f$ can correspond to one of the replicas, i.e., $w=w_i$ and
$f=f_i$. The advantage of MBAR is that we can employ an in principle arbitrary
weight function $w(x)$ (given sufficient statistical weight $e^{-w(x)}$ of the
sampled data) with
\begin{equation}
  \label{eq:app:f}
  f = -\ln \sum_{j=1}^{N_\text{rep}}\sum_{n=1}^{N_\text{t}}
  \frac{e^{-w(x_{jn})}}{N_\text{t}\sum_{k=1}^{N_\text{rep}}(g_j/g_k) 
    e^{f_k-w_k(x_{jn})}}.
\end{equation}
Probabilities $p_i=\mean{\chi_i}$ are calculated as the expectation value of
an indicator function $\chi_i(x)$ which is 1 if the value $x$ falls into bin
$i$ and 0 otherwise. In particular, the distributions shown in
Fig.~\ref{fig:exc} correspond to the unbiased ensemble $P_0[X]$ with $w(x)=0$.

The curves shown in Fig.~\ref{fig:phase} for the mean value $\mean{c}_s$ are
obtained through using the weight function $w(x)=sx$ in
Eq.~\eqref{eq:app:f}. In order to sample trajectories at fixed $s$ we use this
weight function instead of Eq.~\eqref{eq:w} for a chain of replicas with
different $s$ values ranging from $s=0$ to $s=0.01$. To obtain a set of
independent trajectories we keep only every 1000th trajectory for analysis.

\section{Orientational order}
\label{sec:order}

To quantify orientational order we follow Ref.~\cite{stei83}. For each
particle $k$ a complex vector
\begin{equation*}
  q_{lm}(k) \equiv \frac{1}{\Nb}\sum_{k'=1}^{\Nb}
  Y_{lm}(\theta_{kk'},\phi_{kk'})
\end{equation*}
is defined, where $Y_{lm}$ are spherical harmonics and the angles
$\theta_{kk'}$ and $\phi_{kk'}$ describe the orientation of the displacement
vector between particles $k$ and $k'$ with respect to a fixed reference
frame. The sum is over all $\Nb$ neighbors in the first coordination shell
with radius 1.42. The normalized scalar product of the $q$-vectors is
\begin{equation*}
  S(k,k') \equiv \frac{\sum_{m=-6}^6 q_{6m}(k)q^\ast_{6m}(k')}
  {\sqrt{\sum_{m=-6}^6|q_{6m}(k)|^2}\sqrt{\sum_{m=-6}^6|q_{6m}(k')|^2}}.
\end{equation*}
The average over neighbors
\begin{equation}
  \label{eq:psi6}
  \psi_6(k) \equiv \frac{1}{\Nb}\sum_{k'=1}^{\Nb} S(k,k')
\end{equation}
is the bond order parameter. It is $\psi_6=1$ for a particle in a perfect
crystal and acquires a broad distribution with mean $0.2-0.3$ for particles in
a disordered environment.



\begin{thebibliography}{48}
\expandafter\ifx\csname natexlab\endcsname\relax\def\natexlab#1{#1}\fi
\expandafter\ifx\csname bibnamefont\endcsname\relax
  \def\bibnamefont#1{#1}\fi
\expandafter\ifx\csname bibfnamefont\endcsname\relax
  \def\bibfnamefont#1{#1}\fi
\expandafter\ifx\csname citenamefont\endcsname\relax
  \def\citenamefont#1{#1}\fi
\expandafter\ifx\csname url\endcsname\relax
  \def\url#1{\texttt{#1}}\fi
\expandafter\ifx\csname urlprefix\endcsname\relax\def\urlprefix{URL }\fi
\providecommand{\bibinfo}[2]{#2}
\providecommand{\eprint}[2][]{\url{#2}}

\bibitem[{\citenamefont{Angell}(1995)}]{ange95}
\bibinfo{author}{\bibfnamefont{C.~A.} \bibnamefont{Angell}},
  \bibinfo{journal}{Science} \textbf{\bibinfo{volume}{267}},
  \bibinfo{pages}{1924} (\bibinfo{year}{1995}).

\bibitem[{\citenamefont{Debenedetti and Stillinger}(2001)}]{debe01}
\bibinfo{author}{\bibfnamefont{P.~G.} \bibnamefont{Debenedetti}}
  \bibnamefont{and} \bibinfo{author}{\bibfnamefont{F.~H.}
  \bibnamefont{Stillinger}}, \bibinfo{journal}{Nature}
  \textbf{\bibinfo{volume}{410}}, \bibinfo{pages}{259} (\bibinfo{year}{2001}).

\bibitem[{\citenamefont{Berthier et~al.}(2010)\citenamefont{Berthier, Biroli,
  Bouchaud, and Jack}}]{bert10}
\bibinfo{author}{\bibfnamefont{L.}~\bibnamefont{Berthier}},
  \bibinfo{author}{\bibfnamefont{G.}~\bibnamefont{Biroli}},
  \bibinfo{author}{\bibfnamefont{J.-P.} \bibnamefont{Bouchaud}},
  \bibnamefont{and} \bibinfo{author}{\bibfnamefont{R.~L.} \bibnamefont{Jack}},
  \bibinfo{journal}{arXiv:1009.4765}  (\bibinfo{year}{2010}).

\bibitem[{\citenamefont{Weeks et~al.}(2000)\citenamefont{Weeks, Crocker,
  Levitt, Schofield, and Weitz}}]{week00}
\bibinfo{author}{\bibfnamefont{E.~R.} \bibnamefont{Weeks}},
  \bibinfo{author}{\bibfnamefont{J.~C.} \bibnamefont{Crocker}},
  \bibinfo{author}{\bibfnamefont{A.~C.} \bibnamefont{Levitt}},
  \bibinfo{author}{\bibfnamefont{A.}~\bibnamefont{Schofield}},
  \bibnamefont{and} \bibinfo{author}{\bibfnamefont{D.~A.} \bibnamefont{Weitz}},
  \bibinfo{journal}{Science} \textbf{\bibinfo{volume}{287}},
  \bibinfo{pages}{627} (\bibinfo{year}{2000}).

\bibitem[{\citenamefont{Keys et~al.}(2007)\citenamefont{Keys, Abate, Glotzer,
  and Durian}}]{keys07}
\bibinfo{author}{\bibfnamefont{A.~S.} \bibnamefont{Keys}},
  \bibinfo{author}{\bibfnamefont{A.~R.} \bibnamefont{Abate}},
  \bibinfo{author}{\bibfnamefont{S.~C.} \bibnamefont{Glotzer}},
  \bibnamefont{and} \bibinfo{author}{\bibfnamefont{D.~J.}
  \bibnamefont{Durian}}, \bibinfo{journal}{Nature Physics}
  \textbf{\bibinfo{volume}{3}}, \bibinfo{pages}{260} (\bibinfo{year}{2007}).

\bibitem[{\citenamefont{Hansen and McDonald}(2006)}]{hansen}
\bibinfo{author}{\bibfnamefont{J.}~\bibnamefont{Hansen}} \bibnamefont{and}
  \bibinfo{author}{\bibfnamefont{I.}~\bibnamefont{McDonald}},
  \emph{\bibinfo{title}{Theory of Simple Liquids}}
  (\bibinfo{publisher}{Academic Press}, \bibinfo{address}{Amsterdam},
  \bibinfo{year}{2006}), \bibinfo{edition}{3rd} ed.

\bibitem[{\citenamefont{Garrahan and Chandler}(2003)}]{garr03}
\bibinfo{author}{\bibfnamefont{J.~P.} \bibnamefont{Garrahan}} \bibnamefont{and}
  \bibinfo{author}{\bibfnamefont{D.}~\bibnamefont{Chandler}},
  \bibinfo{journal}{Proc. Natl. Acad. Sci. U.S.A.}
  \textbf{\bibinfo{volume}{100}}, \bibinfo{pages}{9710} (\bibinfo{year}{2003}).

\bibitem[{\citenamefont{Chandler and Garrahan}(2010)}]{chan10}
\bibinfo{author}{\bibfnamefont{D.}~\bibnamefont{Chandler}} \bibnamefont{and}
  \bibinfo{author}{\bibfnamefont{J.~P.} \bibnamefont{Garrahan}},
  \bibinfo{journal}{Annu. Rev. Phys. Chem.} \textbf{\bibinfo{volume}{61}},
  \bibinfo{pages}{191} (\bibinfo{year}{2010}).

\bibitem[{\citenamefont{Fredrickson and Andersen}(1984)}]{fred84}
\bibinfo{author}{\bibfnamefont{G.~H.} \bibnamefont{Fredrickson}}
  \bibnamefont{and} \bibinfo{author}{\bibfnamefont{H.~C.}
  \bibnamefont{Andersen}}, \bibinfo{journal}{Phys. Rev. Lett.}
  \textbf{\bibinfo{volume}{53}}, \bibinfo{pages}{1244} (\bibinfo{year}{1984}).

\bibitem[{\citenamefont{Vollmayr-Lee}(2004)}]{voll04}
\bibinfo{author}{\bibfnamefont{K.}~\bibnamefont{Vollmayr-Lee}},
  \bibinfo{journal}{J. Chem. Phys.} \textbf{\bibinfo{volume}{121}},
  \bibinfo{pages}{4781} (\bibinfo{year}{2004}).

\bibitem[{\citenamefont{Widmer-Cooper and Harrowell}(2006)}]{widm06}
\bibinfo{author}{\bibfnamefont{A.}~\bibnamefont{Widmer-Cooper}}
  \bibnamefont{and}
  \bibinfo{author}{\bibfnamefont{P.}~\bibnamefont{Harrowell}},
  \bibinfo{journal}{Phys. Rev. Lett.} \textbf{\bibinfo{volume}{96}},
  \bibinfo{pages}{185701} (\bibinfo{year}{2006}).

\bibitem[{\citenamefont{Widmer-Cooper et~al.}(2009)\citenamefont{Widmer-Cooper,
  Perry, Harrowell, and Reichman}}]{widm09}
\bibinfo{author}{\bibfnamefont{A.}~\bibnamefont{Widmer-Cooper}},
  \bibinfo{author}{\bibfnamefont{H.}~\bibnamefont{Perry}},
  \bibinfo{author}{\bibfnamefont{P.}~\bibnamefont{Harrowell}},
  \bibnamefont{and} \bibinfo{author}{\bibfnamefont{D.~R.}
  \bibnamefont{Reichman}}, \bibinfo{journal}{J. Chem. Phys.}
  \textbf{\bibinfo{volume}{131}}, \bibinfo{pages}{194508}
  (\bibinfo{year}{2009}).

\bibitem[{\citenamefont{Keys et~al.}(2011)\citenamefont{Keys, Hedges, Garrahan,
  Glotzer, and Chandler}}]{keys11}
\bibinfo{author}{\bibfnamefont{A.~S.} \bibnamefont{Keys}},
  \bibinfo{author}{\bibfnamefont{L.~O.} \bibnamefont{Hedges}},
  \bibinfo{author}{\bibfnamefont{J.~P.} \bibnamefont{Garrahan}},
  \bibinfo{author}{\bibfnamefont{S.~C.} \bibnamefont{Glotzer}},
  \bibnamefont{and} \bibinfo{author}{\bibfnamefont{D.}~\bibnamefont{Chandler}},
  \bibinfo{journal}{Phys. Rev. X} \textbf{\bibinfo{volume}{1}},
  \bibinfo{pages}{021013} (\bibinfo{year}{2011}).

\bibitem[{\citenamefont{Merolle et~al.}(2005)\citenamefont{Merolle, Garrahan,
  and Chandler}}]{mero05}
\bibinfo{author}{\bibfnamefont{M.}~\bibnamefont{Merolle}},
  \bibinfo{author}{\bibfnamefont{J.~P.} \bibnamefont{Garrahan}},
  \bibnamefont{and} \bibinfo{author}{\bibfnamefont{D.}~\bibnamefont{Chandler}},
  \bibinfo{journal}{Proc. Natl. Acad. Sci. U.S.A.}
  \textbf{\bibinfo{volume}{102}}, \bibinfo{pages}{10837}
  (\bibinfo{year}{2005}).

\bibitem[{\citenamefont{Jack et~al.}(2006)\citenamefont{Jack, Garrahan, and
  Chandler}}]{jack06a}
\bibinfo{author}{\bibfnamefont{R.~L.} \bibnamefont{Jack}},
  \bibinfo{author}{\bibfnamefont{J.~P.} \bibnamefont{Garrahan}},
  \bibnamefont{and} \bibinfo{author}{\bibfnamefont{D.}~\bibnamefont{Chandler}},
  \bibinfo{journal}{J. Chem. Phys.} \textbf{\bibinfo{volume}{125}},
  \bibinfo{pages}{184509} (\bibinfo{year}{2006}).

\bibitem[{\citenamefont{Garrahan et~al.}(2009)\citenamefont{Garrahan, Jack,
  Lecomte, Pitard, van Duijvendijk, and van Wijland}}]{garr09}
\bibinfo{author}{\bibfnamefont{J.~P.} \bibnamefont{Garrahan}},
  \bibinfo{author}{\bibfnamefont{R.~L.} \bibnamefont{Jack}},
  \bibinfo{author}{\bibfnamefont{V.}~\bibnamefont{Lecomte}},
  \bibinfo{author}{\bibfnamefont{E.}~\bibnamefont{Pitard}},
  \bibinfo{author}{\bibfnamefont{K.}~\bibnamefont{van Duijvendijk}},
  \bibnamefont{and} \bibinfo{author}{\bibfnamefont{F.}~\bibnamefont{van
  Wijland}}, \bibinfo{journal}{J. Phys. A} \textbf{\bibinfo{volume}{42}},
  \bibinfo{pages}{075007} (\bibinfo{year}{2009}).

\bibitem[{\citenamefont{Garrahan et~al.}(2007)\citenamefont{Garrahan, Jack,
  Lecomte, Pitard, van Duijvendijk, and van Wijland}}]{garr07}
\bibinfo{author}{\bibfnamefont{J.~P.} \bibnamefont{Garrahan}},
  \bibinfo{author}{\bibfnamefont{L.}~\bibnamefont{Jack}},
  \bibinfo{author}{\bibfnamefont{V.}~\bibnamefont{Lecomte}},
  \bibinfo{author}{\bibfnamefont{E.}~\bibnamefont{Pitard}},
  \bibinfo{author}{\bibfnamefont{K.}~\bibnamefont{van Duijvendijk}},
  \bibnamefont{and} \bibinfo{author}{\bibfnamefont{F.}~\bibnamefont{van
  Wijland}}, \bibinfo{journal}{Phys. Rev. Lett.} \textbf{\bibinfo{volume}{98}},
  \bibinfo{pages}{195702} (\bibinfo{year}{2007}).

\bibitem[{\citenamefont{Speck and Garrahan}(2011)}]{spec11}
\bibinfo{author}{\bibfnamefont{T.}~\bibnamefont{Speck}} \bibnamefont{and}
  \bibinfo{author}{\bibfnamefont{J.}~\bibnamefont{Garrahan}},
  \bibinfo{journal}{Eur. Phys. J. B} \textbf{\bibinfo{volume}{79}},
  \bibinfo{pages}{1} (\bibinfo{year}{2011}).

\bibitem[{\citenamefont{Jack and Garrahan}(2010)}]{jack10}
\bibinfo{author}{\bibfnamefont{R.~L.} \bibnamefont{Jack}} \bibnamefont{and}
  \bibinfo{author}{\bibfnamefont{J.~P.} \bibnamefont{Garrahan}},
  \bibinfo{journal}{Phys. Rev. E} \textbf{\bibinfo{volume}{81}},
  \bibinfo{pages}{011111} (\bibinfo{year}{2010}).

\bibitem[{\citenamefont{Hedges et~al.}(2009)\citenamefont{Hedges, Jack,
  Garrahan, and Chandler}}]{hedg09}
\bibinfo{author}{\bibfnamefont{L.~O.} \bibnamefont{Hedges}},
  \bibinfo{author}{\bibfnamefont{R.~L.} \bibnamefont{Jack}},
  \bibinfo{author}{\bibfnamefont{J.~P.} \bibnamefont{Garrahan}},
  \bibnamefont{and} \bibinfo{author}{\bibfnamefont{D.}~\bibnamefont{Chandler}},
  \bibinfo{journal}{Science} \textbf{\bibinfo{volume}{323}},
  \bibinfo{pages}{1309} (\bibinfo{year}{2009}).

\bibitem[{\citenamefont{Ritort and Sollich}(2003)}]{rito03}
\bibinfo{author}{\bibfnamefont{F.}~\bibnamefont{Ritort}} \bibnamefont{and}
  \bibinfo{author}{\bibfnamefont{P.}~\bibnamefont{Sollich}},
  \bibinfo{journal}{Adv. Phys.} \textbf{\bibinfo{volume}{52}},
  \bibinfo{pages}{219} (\bibinfo{year}{2003}).

\bibitem[{\citenamefont{J\"ackle and Eisinger}(1991)}]{jack91}
\bibinfo{author}{\bibfnamefont{J.}~\bibnamefont{J\"ackle}} \bibnamefont{and}
  \bibinfo{author}{\bibfnamefont{S.}~\bibnamefont{Eisinger}},
  \bibinfo{journal}{Z. Phys. B} \textbf{\bibinfo{volume}{84}},
  \bibinfo{pages}{115} (\bibinfo{year}{1991}).

\bibitem[{\citenamefont{Kob and Andersen}(1994)}]{kob94}
\bibinfo{author}{\bibfnamefont{W.}~\bibnamefont{Kob}} \bibnamefont{and}
  \bibinfo{author}{\bibfnamefont{H.~C.} \bibnamefont{Andersen}},
  \bibinfo{journal}{Phys. Rev. Lett.} \textbf{\bibinfo{volume}{73}},
  \bibinfo{pages}{1376} (\bibinfo{year}{1994}).

\bibitem[{\citenamefont{Yamamoto and Kob}(2000)}]{yama00}
\bibinfo{author}{\bibfnamefont{R.}~\bibnamefont{Yamamoto}} \bibnamefont{and}
  \bibinfo{author}{\bibfnamefont{W.}~\bibnamefont{Kob}},
  \bibinfo{journal}{Phys. Rev. E} \textbf{\bibinfo{volume}{61}},
  \bibinfo{pages}{5473} (\bibinfo{year}{2000}).

\bibitem[{\citenamefont{Berthier and Tarjus}(2010)}]{bert10a}
\bibinfo{author}{\bibfnamefont{L.}~\bibnamefont{Berthier}} \bibnamefont{and}
  \bibinfo{author}{\bibfnamefont{G.}~\bibnamefont{Tarjus}},
  \bibinfo{journal}{Phys. Rev. E} \textbf{\bibinfo{volume}{82}},
  \bibinfo{pages}{031502} (\bibinfo{year}{2010}).

\bibitem[{\citenamefont{Karmakar et~al.}(2009)\citenamefont{Karmakar, Dasgupta,
  and Sastry}}]{karm09}
\bibinfo{author}{\bibfnamefont{S.}~\bibnamefont{Karmakar}},
  \bibinfo{author}{\bibfnamefont{C.}~\bibnamefont{Dasgupta}}, \bibnamefont{and}
  \bibinfo{author}{\bibfnamefont{S.}~\bibnamefont{Sastry}},
  \bibinfo{journal}{Proc. Natl. Acad. Sci. U.S.A.}
  \textbf{\bibinfo{volume}{106}}, \bibinfo{pages}{3675} (\bibinfo{year}{2009}).

\bibitem[{\citenamefont{Stillinger and Weber}(1984)}]{stil84}
\bibinfo{author}{\bibfnamefont{F.}~\bibnamefont{Stillinger}} \bibnamefont{and}
  \bibinfo{author}{\bibfnamefont{T.}~\bibnamefont{Weber}},
  \bibinfo{journal}{Science} \textbf{\bibinfo{volume}{225}},
  \bibinfo{pages}{983} (\bibinfo{year}{1984}).

\bibitem[{\citenamefont{Elmatad et~al.}(2009)\citenamefont{Elmatad, Chandler,
  and Garrahan}}]{elma09}
\bibinfo{author}{\bibfnamefont{Y.~S.} \bibnamefont{Elmatad}},
  \bibinfo{author}{\bibfnamefont{D.}~\bibnamefont{Chandler}}, \bibnamefont{and}
  \bibinfo{author}{\bibfnamefont{J.~P.} \bibnamefont{Garrahan}},
  \bibinfo{journal}{J. Phys. Chem. B} \textbf{\bibinfo{volume}{113}},
  \bibinfo{pages}{5563–5567} (\bibinfo{year}{2009}).

\bibitem[{\citenamefont{Chandler}(1987)}]{chandler}
\bibinfo{author}{\bibfnamefont{D.}~\bibnamefont{Chandler}},
  \emph{\bibinfo{title}{Introduction to Modern Statistical Mechanics}}
  (\bibinfo{publisher}{Oxford University Press}, \bibinfo{address}{Oxford},
  \bibinfo{year}{1987}).

\bibitem[{\citenamefont{Widmer-Cooper et~al.}(2004)\citenamefont{Widmer-Cooper,
  Harrowell, and Fynewever}}]{widm04}
\bibinfo{author}{\bibfnamefont{A.}~\bibnamefont{Widmer-Cooper}},
  \bibinfo{author}{\bibfnamefont{P.}~\bibnamefont{Harrowell}},
  \bibnamefont{and}
  \bibinfo{author}{\bibfnamefont{H.}~\bibnamefont{Fynewever}},
  \bibinfo{journal}{Phys. Rev. Lett.} \textbf{\bibinfo{volume}{93}},
  \bibinfo{pages}{135701} (\bibinfo{year}{2004}).

\bibitem[{\citenamefont{Fern\'andez and Harrowell}(2004)}]{fern04}
\bibinfo{author}{\bibfnamefont{J.~R.} \bibnamefont{Fern\'andez}}
  \bibnamefont{and}
  \bibinfo{author}{\bibfnamefont{P.}~\bibnamefont{Harrowell}},
  \bibinfo{journal}{J. Phys. Chem. B} \textbf{\bibinfo{volume}{108}},
  \bibinfo{pages}{6850} (\bibinfo{year}{2004}).

\bibitem[{\citenamefont{Coslovich and Pastore}(2007)}]{cosl07}
\bibinfo{author}{\bibfnamefont{D.}~\bibnamefont{Coslovich}} \bibnamefont{and}
  \bibinfo{author}{\bibfnamefont{G.}~\bibnamefont{Pastore}},
  \bibinfo{journal}{J. Chem. Phys.} \textbf{\bibinfo{volume}{127}},
  \bibinfo{pages}{124504} (\bibinfo{year}{2007}).

\bibitem[{\citenamefont{Pedersen et~al.}(2010)\citenamefont{Pedersen, Schroder,
  Dyre, and Harrowell}}]{pede10}
\bibinfo{author}{\bibfnamefont{U.~R.} \bibnamefont{Pedersen}},
  \bibinfo{author}{\bibfnamefont{T.~B.} \bibnamefont{Schroder}},
  \bibinfo{author}{\bibfnamefont{J.~C.} \bibnamefont{Dyre}}, \bibnamefont{and}
  \bibinfo{author}{\bibfnamefont{P.}~\bibnamefont{Harrowell}},
  \bibinfo{journal}{Phys. Rev. Lett.} \textbf{\bibinfo{volume}{104}},
  \bibinfo{pages}{105701} (\bibinfo{year}{2010}).

\bibitem[{\citenamefont{Vogel and Glotzer}(2004)}]{voge04}
\bibinfo{author}{\bibfnamefont{M.}~\bibnamefont{Vogel}} \bibnamefont{and}
  \bibinfo{author}{\bibfnamefont{S.~C.} \bibnamefont{Glotzer}},
  \bibinfo{journal}{Phys. Rev. Lett.} \textbf{\bibinfo{volume}{92}},
  \bibinfo{pages}{255901} (\bibinfo{year}{2004}).

\bibitem[{\citenamefont{Heuer}(2008)}]{heue08}
\bibinfo{author}{\bibfnamefont{A.}~\bibnamefont{Heuer}}, \bibinfo{journal}{J.
  Phys.: Condens. Matter} \textbf{\bibinfo{volume}{20}},
  \bibinfo{pages}{373101} (\bibinfo{year}{2008}).

\bibitem[{\citenamefont{Goldstein}(1969)}]{gold69}
\bibinfo{author}{\bibfnamefont{M.}~\bibnamefont{Goldstein}},
  \bibinfo{journal}{J. Chem. Phys.} \textbf{\bibinfo{volume}{51}},
  \bibinfo{pages}{3728} (\bibinfo{year}{1969}).

\bibitem[{\citenamefont{Fris et~al.}(2011)\citenamefont{Fris, Appignanesi, and
  Weeks}}]{fris11}
\bibinfo{author}{\bibfnamefont{J.~A.~R.} \bibnamefont{Fris}},
  \bibinfo{author}{\bibfnamefont{G.~A.} \bibnamefont{Appignanesi}},
  \bibnamefont{and} \bibinfo{author}{\bibfnamefont{E.~R.} \bibnamefont{Weeks}},
  \bibinfo{journal}{Phys. Rev. Lett.} \textbf{\bibinfo{volume}{107}},
  \bibinfo{pages}{065704} (\bibinfo{year}{2011}).

\bibitem[{\citenamefont{Gebremichael et~al.}(2004)\citenamefont{Gebremichael,
  Vogel, and Glotzer}}]{gebr04}
\bibinfo{author}{\bibfnamefont{Y.}~\bibnamefont{Gebremichael}},
  \bibinfo{author}{\bibfnamefont{M.}~\bibnamefont{Vogel}}, \bibnamefont{and}
  \bibinfo{author}{\bibfnamefont{S.~C.} \bibnamefont{Glotzer}},
  \bibinfo{journal}{J. Chem. Phys.} \textbf{\bibinfo{volume}{120}},
  \bibinfo{pages}{4415} (\bibinfo{year}{2004}).

\bibitem[{\citenamefont{Chen et~al.}(2011)\citenamefont{Chen, Manning, Yunker,
  Ellenbroek, Zhang, Liu, and Yodh}}]{chen11}
\bibinfo{author}{\bibfnamefont{K.}~\bibnamefont{Chen}},
  \bibinfo{author}{\bibfnamefont{M.~L.} \bibnamefont{Manning}},
  \bibinfo{author}{\bibfnamefont{P.~J.} \bibnamefont{Yunker}},
  \bibinfo{author}{\bibfnamefont{W.~G.} \bibnamefont{Ellenbroek}},
  \bibinfo{author}{\bibfnamefont{Z.}~\bibnamefont{Zhang}},
  \bibinfo{author}{\bibfnamefont{A.~J.} \bibnamefont{Liu}}, \bibnamefont{and}
  \bibinfo{author}{\bibfnamefont{A.~G.} \bibnamefont{Yodh}},
  \bibinfo{journal}{Phys. Rev. Lett.} \textbf{\bibinfo{volume}{107}},
  \bibinfo{pages}{108301} (\bibinfo{year}{2011}).

\bibitem[{\citenamefont{Jack et~al.}(2011)\citenamefont{Jack, Hedges, Garrahan,
  and Chandler}}]{jack11}
\bibinfo{author}{\bibfnamefont{R.~L.} \bibnamefont{Jack}},
  \bibinfo{author}{\bibfnamefont{L.~O.} \bibnamefont{Hedges}},
  \bibinfo{author}{\bibfnamefont{J.~P.} \bibnamefont{Garrahan}},
  \bibnamefont{and} \bibinfo{author}{\bibfnamefont{D.}~\bibnamefont{Chandler}},
  \bibinfo{journal}{Phys. Rev. Lett.} \textbf{\bibinfo{volume}{107}},
  \bibinfo{pages}{275702} (\bibinfo{year}{2011}).

\bibitem[{\citenamefont{Plimpton}(1995)}]{lammps}
\bibinfo{author}{\bibfnamefont{S.}~\bibnamefont{Plimpton}},
  \bibinfo{journal}{J. Comp. Phys.} \textbf{\bibinfo{volume}{117}},
  \bibinfo{pages}{1} (\bibinfo{year}{1995}), \bibinfo{note}{available at
  http://lammps.sandia.gov}.

\bibitem[{\citenamefont{Bitzek et~al.}(2006)\citenamefont{Bitzek, Koskinen,
  G\"ahler, Moseler, and Gumbsch}}]{bitz06}
\bibinfo{author}{\bibfnamefont{E.}~\bibnamefont{Bitzek}},
  \bibinfo{author}{\bibfnamefont{P.}~\bibnamefont{Koskinen}},
  \bibinfo{author}{\bibfnamefont{F.}~\bibnamefont{G\"ahler}},
  \bibinfo{author}{\bibfnamefont{M.}~\bibnamefont{Moseler}}, \bibnamefont{and}
  \bibinfo{author}{\bibfnamefont{P.}~\bibnamefont{Gumbsch}},
  \bibinfo{journal}{Phys. Rev. Lett.} \textbf{\bibinfo{volume}{97}},
  \bibinfo{pages}{170201} (\bibinfo{year}{2006}).

\bibitem[{\citenamefont{Frenkel and Smit}(2002)}]{frenkel}
\bibinfo{author}{\bibfnamefont{D.}~\bibnamefont{Frenkel}} \bibnamefont{and}
  \bibinfo{author}{\bibfnamefont{B.}~\bibnamefont{Smit}},
  \emph{\bibinfo{title}{Understanding Molecular Simulation: From Algorithms to
  Applications}} (\bibinfo{publisher}{Academic Press}, \bibinfo{address}{San
  Diego}, \bibinfo{year}{2002}), \bibinfo{edition}{2nd} ed.

\bibitem[{\citenamefont{Bolhuis et~al.}(2002)\citenamefont{Bolhuis, Chandler,
  Dellago, and Geissler}}]{bolh02}
\bibinfo{author}{\bibfnamefont{P.~G.} \bibnamefont{Bolhuis}},
  \bibinfo{author}{\bibfnamefont{D.}~\bibnamefont{Chandler}},
  \bibinfo{author}{\bibfnamefont{C.}~\bibnamefont{Dellago}}, \bibnamefont{and}
  \bibinfo{author}{\bibfnamefont{P.~L.} \bibnamefont{Geissler}},
  \bibinfo{journal}{Annu. Rev. Phys. Chem.} \textbf{\bibinfo{volume}{53}},
  \bibinfo{pages}{291} (\bibinfo{year}{2002}).

\bibitem[{\citenamefont{Dellago et~al.}(2002)\citenamefont{Dellago, Bolhuis,
  and Geissler}}]{dell02}
\bibinfo{author}{\bibfnamefont{C.}~\bibnamefont{Dellago}},
  \bibinfo{author}{\bibfnamefont{P.~G.} \bibnamefont{Bolhuis}},
  \bibnamefont{and} \bibinfo{author}{\bibfnamefont{P.~L.}
  \bibnamefont{Geissler}}, \bibinfo{journal}{Adv. Chem. Phys.}
  \textbf{\bibinfo{volume}{123}}, \bibinfo{pages}{1} (\bibinfo{year}{2002}).

\bibitem[{\citenamefont{Shirts and Chodera}(2008)}]{shir08}
\bibinfo{author}{\bibfnamefont{M.~R.} \bibnamefont{Shirts}} \bibnamefont{and}
  \bibinfo{author}{\bibfnamefont{J.~D.} \bibnamefont{Chodera}},
  \bibinfo{journal}{J. Chem. Phys.} \textbf{\bibinfo{volume}{129}},
  \bibinfo{pages}{124105} (\bibinfo{year}{2008}).

\bibitem[{\citenamefont{Minh and Chodera}(2009)}]{minh09}
\bibinfo{author}{\bibfnamefont{D.~D.~L.} \bibnamefont{Minh}} \bibnamefont{and}
  \bibinfo{author}{\bibfnamefont{J.~D.} \bibnamefont{Chodera}},
  \bibinfo{journal}{J. Chem. Phys.} \textbf{\bibinfo{volume}{131}},
  \bibinfo{pages}{134110} (\bibinfo{year}{2009}).

\bibitem[{\citenamefont{Steinhardt et~al.}(1983)\citenamefont{Steinhardt,
  Nelson, and Ronchetti}}]{stei83}
\bibinfo{author}{\bibfnamefont{P.~J.} \bibnamefont{Steinhardt}},
  \bibinfo{author}{\bibfnamefont{D.~R.} \bibnamefont{Nelson}},
  \bibnamefont{and}
  \bibinfo{author}{\bibfnamefont{M.}~\bibnamefont{Ronchetti}},
  \bibinfo{journal}{Phys. Rev. B} \textbf{\bibinfo{volume}{28}},
  \bibinfo{pages}{784} (\bibinfo{year}{1983}).

\end{thebibliography}
\end{document}